\documentclass[3p,times,twocolumn,fleqn]{elsarticle}
 \biboptions{comma,sort&compress}
 
\usepackage{here}
\usepackage{graphicx}
\usepackage{dsfont}
\usepackage[fleqn]{amsmath}

\usepackage{ecrc}

\volume{00}

\firstpage{1}

\journalname{Nuclear and Particle Physics Proceedings}

\runauth{}

\jid{nppp}

\jnltitlelogo{Nuclear and Particle Physics Proceedings}

\usepackage{amssymb}

\usepackage[figuresright]{rotating}

\begin{document}

\begin{frontmatter}

\title{The pion in holographic light-front QCD$^*$}
 \cortext[cor0]{Talk given at 20th International Conference in Quantum Chromodynamics (QCD 17),  3 July - 7 July 2017, Montpellier - FR}
 \author[label2]{M. Ahmady}
\ead{mahamdy@mta.ca}

 \author[label3]{F. Chishtie\corref{cor1}}
  \cortext[cor1]{Now at the Asian Disaster Preparedness Center, Thailand.}
\ead{fchistie@uwo.ca}
\address[label3]{Department of Physics and Astronomy, University of Western Ontario, Ontario, Canada}
 \author[label1,label2]{R. Sandapen\fnref{fn1}}
 \address[label1]{Physics Department, Acadia University, Wolfville, Nova Scotia B4P 2R6, Canada}
\address[label2]{Physics Department, Mount Allison University, Sackville, New Brunswick E4L 1E6, Canada}   \fntext[fn1]{Speaker, Corresponding author.}
    \ead{ruben.sandapen@acadiau.ca}

\pagestyle{myheadings}
\markright{ }
\begin{abstract}
We report that the pion radius, decay constant, EM and transition form factors can simultaneously and accurately be predicted in light-front holographic QCD with a universal AdS/QCD mass scale, provided dynamical spin effects are taken into account. 
\end{abstract}

\end{frontmatter}
\section{Light-front holographic QCD}
The ultimate goal in light-front QCD is to solve the Heisenberg eigenvalue problem:
\begin{flalign}
H_{\mathrm{LF}}^{\mathrm{QCD}} |\Psi(P) \rangle=M^2 |\Psi(P) \rangle
\label{Heisenberg}
\end{flalign}
where $H_{\mathrm{LF}}^{\mathrm{QCD}} \equiv H= P^-P^+ -\mathbf{P}_{\perp}^2$ is the QCD light-front Hamiltonian,\footnote{The QCD light-front Hamiltonian is given explicitly in \cite{Brodsky:1997de}.} $P^\mu$ is the hadron momentum and $P^{\mu} P_{\mu}=M^2$ is the hadron mass squared. To make contact with the constituent quark model (CQM), one can recast Eq. \ref{Heisenberg} as an eigenvalue problem for the valence Fock sector acted upon by an effective Hamiltonian which now includes the effects of the higher Fock sectors on the valence sector as well as retarded interactions \cite{Brodsky:1997de}. Doing so, we have
 \begin{equation}
	H_{\mathrm{eff}} \mathcal{P}|\Psi \rangle =M^2 \mathcal{P}|\Psi \rangle 
\label{effLFSE}
\end{equation}
where 
\begin{equation}
	H_{\mathrm{eff}}=\mathcal{P} H \mathcal{P} + \mathcal{P}H\mathcal{Q} \left(\frac{1}{M^2-\mathcal{Q}H \mathcal{Q}} \right) \mathcal{Q} H \mathcal{P} 
\label{Heff}
\end{equation}
with $\mathcal{P}$ being the valence Fock sector projection operator and $\mathcal{Q}=\mathds{1}-\mathcal{P}$ its complement. Using the fact that $H_{\mathrm{LF}}^{\mathrm{QCD}}=T+U$, where $T$ the kinetic energy operator, which is diagonal in Fock space, Eq. \ref{Heff} becomes
\begin{equation}
	H_{\mathrm{eff}}=\mathcal{P} T \mathcal{P} + \mathcal{P} U_{\mathrm{eff}} \mathcal{P} 
	\end{equation}
where the effective potential is given by
\begin{equation}
	\mathcal{P} U_{\mathrm{eff}} \mathcal{P} = \mathcal{P} U \mathcal{P} + \mathcal{P} U\mathcal{Q} \left(\frac{1}{M^2-\mathcal{Q} H \mathcal{Q}} \right) \mathcal{Q} U \mathcal{P} \;.
	\end{equation}
The second term in $U_{\mathrm{eff}}$ accounts for the coupling of the valence sector to higher Fock sectors as well as for retarded interactions (since it contains the unknown eigenvalue $M^2$). Its exact computation in QCD is a formidable task, as difficult as solving Eq. \ref{Heisenberg} itself. Projecting out the valence light-front wavefunction from Eq. \ref{effLFSE}, we obtain
\begin{flalign}
	\left(\frac{\mathbf{k}_{\perp}^2 + m^2}{x(1-x)} + U_{\mathrm{eff}}(x,\mathbf{k}_{\perp}) \right) \Psi(x,\mathbf{k_{\perp}},\lambda,\lambda^{\prime})\nonumber \\ = M^2 \Psi(x,\mathbf{k_{\perp}},\lambda,\lambda^{\prime})  
	\label{LFSE}
\end{flalign}	
where $\Psi(x,\mathbf{k_{\perp}},\lambda, \lambda^{\prime})=\langle q \bar{q}:x,\mathbf{k_{\perp}},\lambda,\lambda^{\prime}|\Psi\rangle$ is the valence wavefunction, with $x$ being the light-front momentum fraction carried by the quark of mass $m$, $\mathbf{k_{\perp}}$ its transverse momentum and $\lambda$ its helicity ($\lambda^{\prime}$ is the helicity of the antiquark).  


Although the exact form of $U_{\mathrm{eff}}(x,\mathbf{k_{\perp}})$ is unknown, some insights into its analytic form can be found in the so-called semiclassical approximation \cite{Brodsky:2014yha} whereby quantum loops as well as quark masses are neglected. In this approximation, the effective potential is assumed to be a function of the $q\bar{q}$ invariant mass squared evaluated in the limit of vanishing quark masses: $U_{\mathrm{eff}}(x,\mathbf{k_{\perp}}) \to U_{\mathrm{eff}}(\mathcal{M}^2|_{m \to 0})$ 
where
\begin{equation}
	\mathcal{M}^2|_{m \to 0} = \frac{\mathbf{k}_{\perp}^2}{x(1-x)} \;.
\end{equation}
We then proceed by suppressing the helicity indices of the light-front wavefunction \cite{deTeramond:2005su,Brodsky:2006uqa,deTeramond:2008ht,Brodsky:2014yha}. Note that this is only justified if the helicity dependence decouples from the dynamics, i.e.
\begin{equation}
	\Psi(x,\mathbf{k_{\perp}}, \lambda, \lambda^\prime) \to  \Psi(x,\mathbf{k_{\perp}}) S_{\lambda \lambda^\prime}
	\end{equation}
	where, for the pion, $S_{\lambda \lambda^\prime}=\frac{1}{\sqrt{2}}\lambda \delta_{\lambda,-\lambda^\prime}$. Then, introducing the 2-dimensional Fourier transform of $\mathcal{M}^2|_{m \to 0}$,
	\begin{equation}
	\mathbf{\zeta}^2 = x(1-x) \mathbf{b}^2_{\perp}
	\end{equation}
	allows a separation of variables:
	\begin{equation}
	\Psi(x,\zeta,\varphi)=\frac{\phi(\zeta)}{\sqrt{2\pi \zeta}} X(x) e^{iL\varphi} \;,
	\end{equation}
	so that Eq. \ref{LFSE} leads to
	\begin{equation}
			\left(-\frac{\mathrm{d}^2}{\mathrm{d}\zeta^2}-\frac{1-4L^2}{4\zeta^2} + U_{\mathrm{eff}}(\zeta) \right) \phi(\zeta)=M^2 \phi(\zeta) \;.
	\label{hSE}
	\end{equation}
	We refer to Eq. \ref{hSE} as the holographic Schr\"odinger Equation. This is because with $\zeta \leftrightarrow z$, where $z$ is the $5^{\mathrm{th}}$ dimension in anti-de Sitter space, and $L^2 \leftrightarrow (\mu R)^2 + (2-J)^2$ where $\mu$ is a 5-dimensional mass parameter and $R$ is the radius of curvature of the AdS space, Eq. \ref{hSE} maps onto the classical wave equation for weakly-coupled, propagating spin-$J$ string modes in AdS  \cite{deTeramond:2005su,Brodsky:2006uqa,deTeramond:2008ht,Brodsky:2007hb}. The confining QCD potential is then determined by the form of the dilaton field which breaks conformal invariance in AdS: \cite{Brodsky:2014yha}
	\begin{equation}
	U_{\mathrm{eff}}(\zeta)= \frac{1}{2} \varphi^{\prime \prime}(z) + \frac{1}{4} \varphi^{\prime}(z)^2 + \frac{2J-3}{2 z} \varphi^{\prime}(z) \;.
	\label{dilaton-potential}
	\end{equation}

	At this point, the form of the dilaton field is unspecified. However, it turns out that its form can be uniquely determined \cite{Brodsky:2013ar}. To see this, recall that in the semiclassical approximation, we are neglecting quark masses and quantum loops. Neglecting quantum loops means that we are neglecting short-distance effects and consequently no QCD scale, $\Lambda_{\mathrm{QCD}}$, appears via dimensional transmutation. Neglecting quark masses means that we are in the chiral limit, $m \to 0$ of QCD. In the absence of $\Lambda_{\mathrm{QCD}}$, the chiral limit requires that $m \ll \kappa$ where $\kappa$ is a mass scale which is yet to emerge and which sets both the hadronic mass and confinement scales. A hint as to how the mass scale $\kappa$ can emerge can be found in the earlier investigations of de Alfaro, Furbini and Furlan (dAFF) on conformal quantum mechanics \cite{deAlfaro:1976vlx}. dAFF showed that it is possible, via a change in evolution parameter, to introduce a mass scale in an otherwise conformally invariant quantum mechanical Hamiltonian while still preserving the conformal symmetry of the underlying action. It then follows that the conformal-symmetry breaking term in the new Hamiltonian is quadratic in the generalized coordinate. Applying the dAFF mechanism to the holographic light-front Hamiltonian,  it follows that the confining potential must be that of a harmonic oscillator: $U^{\mathrm{dAFF}}_{\mathrm{eff}}=\kappa^4 \zeta^2$. To recover this harmonic potential, the dilaton field has to be quadratic: $\varphi(z)=\kappa^2 z^2$, which coincides with the phenomenological soft-wall model of \cite{Karch:2006pv}. Eq. \ref{dilaton-potential} then implies that 
	\begin{equation}
		U_{\mathrm{eff}}(\zeta)=\kappa^4 \zeta^2 + 2 \kappa^2 (J-1)	
	\label{hUeff}	
	\end{equation}
	where $J=L+S$. 
	
	
 Solving the holographic Schr\"odinger Equation with the confining potential given by Eq. \ref{hUeff} yields the mass spectrum
\begin{equation}
 	M^2= 4\kappa^2 \left(n+L +\frac{S}{2}\right)\;
 	\label{mass-Regge}
 \end{equation}
and the wavefunctions
 \begin{flalign}
 	\phi_{nL}(\zeta)= \kappa^{1+L} \sqrt{\frac{2 n !}{(n+L)!}} \zeta^{1/2+L} \exp{\left(\frac{-\kappa^2 \zeta^2}{2}\right)} \nonumber \\ \times ~ L_n^L(\kappa^2 \zeta^2)\;.
 \label{phi-zeta}
 \end{flalign}
The immediate striking prediction is that the lowest lying bound state, with quantum numbers $n=L=S=0$, is massless: $M^2=0$.  This state is naturally identified with the pion since general chiral symmetry arguments require that the pion mass vanishes in chiral limit $m \to 0$. Note that the harmonic oscillator confining potential uniquely leads to a massless pion \cite{Brodsky:2013npa}: if we consider a more general potential $U_{\mathrm{eff}}(\zeta) \propto \zeta^p$, we obtain $M_{\pi}=0$ only for $p=2$. 
 
 To completely specify the light-front wavefunction, we need to fix $X(x)$. This is done by matching the EM form (or gravitational) factor  in physical spacetime and AdS, resulting in $X(x)=\sqrt{x(1-x)}$ \cite{Brodsky:2008pf}. The normalized holographic light-front wavefunction for the pion is then
 \begin{equation}
 	\Psi^{\pi} (x,\zeta^2) = \frac{\kappa}{\sqrt{\pi}} \sqrt{x (1-x)}  \exp{ \left[ -{ \kappa^2 \zeta^2  \over 2} \right] } \;.
\label{pionhwf} 
\end{equation}
The generalization of Eq. \ref{pionhwf} for non-vanishing light quark masses is straightforward \cite{Brodsky:2014yha}
and leads to 
\begin{flalign}
\Psi^{\pi} (x,\zeta^2) = \mathcal{N} \sqrt{x (1-x)}  \exp{ \left[ -{ \kappa^2 \zeta^2  \over 2} \right] } \nonumber \\
\times \exp{ \left[ -{m_f^2 \over 2 \kappa^2 x(1-x) } \right]}
\end{flalign}
where $\mathcal{N}$ is a normalization constant which is fixed by requiring that 
 \begin{equation}
 	\int \mathrm{d}^2 \mathbf{b} \mathrm{d} x |\Psi^{\pi}(x,\zeta^2)|^2 = P_{q\bar{q}} 
 	\label{norm}
 \end{equation}
with $P_{q\bar{q}}$ being the probability of finding the pion in the valence Fock sector.

\section{Dynamical spin effects}
It remains to fix the value of the AdS/QCD mass scale $\kappa$. Since the pion mass vanishes for any $\kappa$, one can use instead the $\rho$ mass to fix $\kappa$, i.e. $\kappa=M_{\rho}/\sqrt{2}=540$ MeV \cite{Forshaw:2012im}. More generally, $\kappa$ can be adjusted to fit the slopes of the Regge trajectories predicted by Eq. \ref{mass-Regge}. Ref. \cite{Brodsky:2014yha} reports $\kappa=590$ MeV for pseudoscalar mesons and $\kappa=540$ MeV for vector mesons while a recent fit to the  Regge slopes of mesons and baryons, treated as conformal superpartners, yields  $\kappa=523$ MeV \cite{Brodsky:2016rvj}. With $\kappa=523$ MeV and the $\beta$-function of the QCD running coupling at $5$-loops,  Brodsky, Deur and de T\'eramond recently predicted the QCD renormalization scale, $\Lambda^{{\overline{MS}}}_{\mathrm{QCD}}$, in excellent agreement with the world average value \cite{Deur:2016opc}. Furthermore, $\kappa=540-560$ MeV leads to successful predictions for  diffractive light vector meson production \cite{Forshaw:2012im,Ahmady:2016ujw}.  These findings hint towards the emergence of a  universal fundamental AdS/QCD scale $\kappa \sim 500$ MeV.  

\begin{table*}[hbt]
\setlength{\tabcolsep}{1.5pc}
\newlength{\digitwidth} \settowidth{\digitwidth}{\rm 0}
\catcode`?=\active \def?{\kern\digitwidth}
\caption{Parameters in previous pion work}
\label{tab:pionparams}
\begin{tabular*}{\textwidth}{@{}l@{\extracolsep{\fill}}rrrr}
\hline
Reference & $\kappa$ [MeV] & $m_{u/d}$ [MeV] & $P_{q\bar{q}}$ \\
\hline
Brodsky et al. \cite{Brodsky:2011xx} &$432$ & $0$ & $0.5$\\

Vega et al.  \cite{Vega:2009zb}   & $787$ & $330$ & $0.279$ \\
Branz et al. \cite{Branz:2010ub}    & $550$ & $420$ & $0.6$ \\
Swarnkar $\&$ Chakrabarti \cite{Swarnkar:2015osa}           & $550$ & $330$ & $0.61$\\
\hline
\end{tabular*}
\end{table*}

In earlier applications of light-front holography with massless quarks, much lower values of $\kappa$ were required to fit the pion data: $\kappa=375$ MeV in Ref. \cite{Brodsky:2007hb} in order to fit the pion EM form factor data and $\kappa=432$ MeV (with $P_{q\bar{q}}=0.5$) to fit the photon-to-pion transition form factor data simultaneously at large $Q^2$ and $Q^2=0$  \cite{Brodsky:2011xx}. In more recent work using constituent quark masses \cite{Vega:2008te,Swarnkar:2015osa,Vega:2009zb,Branz:2010ub}, it seems that a universal value of $\kappa$ can be used only if the assumption that the pion consists only of the leading valence Fock sector is relaxed:  $P_{q\bar{q}} < 1$, as can be seen in Table \ref{tab:pionparams}. Note that this sets the pion as a special case, since for other mesons $P_{q\bar{q}}=1$ \cite{Branz:2010ub,Swarnkar:2015osa}. 

As explained in \cite{Ahmady:2016ufq}, we find that it is not necessary to treat the pion as a special case if dynamical spin effects are taken into account as was previously done in \cite{Forshaw:2012im,Ahmady:2016ujw} for light vector mesons. We restore the helicity dependence by assuming that 
\begin{equation}
	\Psi(x,\mathbf{k_{\perp}}) \to \Psi(x, \mathbf{k_{\perp}}) S_{\lambda \lambda^{\prime}} (x, \mathbf{k_{\perp}})	
	\label{spin-improved-wf}
	\end{equation}
	where, for the pion, we choose
		\begin{align}
	S_{\lambda \lambda^{\prime}} (x, \mathbf{k_{\perp}})= \frac{\bar{v}_{\lambda^{\prime}}(x,\mathbf{k_{\perp}})}{\sqrt{1-x}} \left[A \frac{M_{\pi}^2}{P^+} \gamma^+ \gamma^5 + B M_{\pi} \gamma^5 \right] \nonumber \\ \times \frac{u_{\lambda}(x,\mathbf{k_{\perp}})}{\sqrt{x}} 
	\label{spin-structure} 
	\end{align}
	where $A$ and $B$ are constants. Eq. \ref{spin-structure} is a general form for the coupling of a point-like pseudoscalar particle to fermions. By using Eq. \ref{spin-improved-wf}, we are assuming that light-front perturbation theory determines the spin structure of the pion and consequently that all bound state effects are fully captured by the holographic pion wavefunction $\Psi(x, \mathbf{k_{\perp}})$. 
	
	We consider three cases: $[A=1,B=0];[A=0,B=1];[A=B=1]$, noting that the first case leads to a decoupling of the quarks' spins from their dynamics i.e. with $[A=1,B=0]$, we recover the original holographic wavefunction without dynamical spin effects. The generalization of the normalization condition, Eq. \ref{norm}, for our spin-improved holographic wavefunction is:
	\begin{equation}
		\sum_{\lambda, \lambda^{\prime}} \int \mathrm{d}^2 \mathbf{b} \mathrm{d} x |\Psi^{\pi}(x,\zeta^2,\lambda,\lambda^{\prime})|^2 = 1 \;. 
 	\label{newnorm}
    \end{equation}
	Note that we have taken $P_{q\bar{q}}=1$, i.e. we assume that the valence Fock sector dominates over higher Fock sectors, thus enforcing a CQM picture of the pion. Note that this does not necessarily imply that the effects of higher Fock sectors in the valence sector are also negligible. The latter are, in principle,  included in the effective potential $U_{\mathrm{eff}}$ and consequently as argued in \cite{Brodsky:2014yha}, the quark masses appearing in Eq. \ref{LFSE} should be interpreted as effective quark masses, resulting from the coupling of the valence sector with higher Fock sectors.

\section{Predictions}
\begin{figure}[htb]
\includegraphics[width=1.0\linewidth]{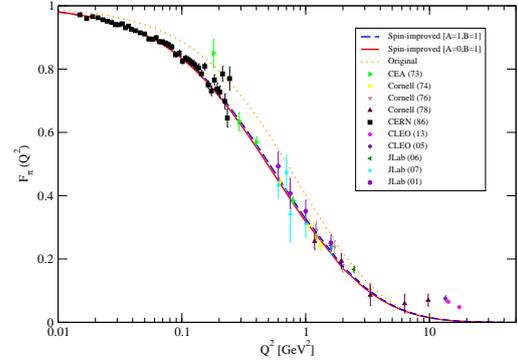}
\caption{Predictions for the pion EM form factor using our spin-improved holographic wavefunction (solid red and dashed black) and the original holographic wavefunction (dotted orange). The references for the data can be found in \cite{Ahmady:2016ufq}.}
\label{fig:EMFF}
\end{figure}

\begin{figure}[htb]
\includegraphics[width=0.9\linewidth]{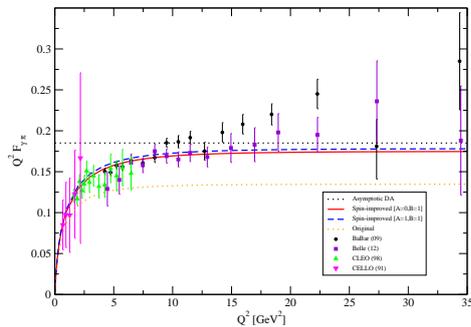}
\caption{Predictions for the photon-to-pion transition form factor using our spin-improved holographic wavefunction (solid red and dashed black) and the original holographic wavefunction (dotted orange). The references for the data can be found in \cite{Ahmady:2016ufq}.}
\label{fig:TFF}
\end{figure}

	Having specified our spin-improved pion holographic light-front wavefunction, we now proceed to make predictions for the pion radius, EM form factor, decay constant and transition form factor. To generate all predictions, we choose to use a constituent quark mass $m_{u/d}=330$ MeV. The pion charge radius and decay constant are given by \cite{Ahmady:2016ufq}
\begin{equation}
	\sqrt{\langle r_{\pi}^2 \rangle} = \left[\frac{3}{2} \int \mathrm{d} x \mathrm{d}^2 \mathbf{b} [b (1-x)]^2 |\Psi^{\pi}(x,\mathbf{b})|^2 \right]^{1/2}	
	\label{radius}
	\end{equation}
and 
\begin{flalign}
	f_{\pi}= 2 \sqrt{\frac{N_c}{\pi}}  \int \mathrm{d} x   \{A((x(1-x) M_{\pi}^2)+ B m_f M_{\pi}\} \nonumber \\ \left. \times \frac{\Psi^{\pi} (x,\zeta)}{x(1-x)}\right|_{\zeta=0}	
\label{decayconstant}
\end{flalign}
respectively while the EM and transition pion form factors are given by 
\begin{equation}
F_{\pi}(Q^2)= 2 \pi \int \mathrm{d} x \mathrm{d} b ~ b ~ J_{0}[(1-x)  b Q] ~ |\Psi^{\pi}(x,\textbf{b})|^2
\label{EMFF}
\end{equation}
and 
\begin{equation}
F_{\gamma \pi} (Q^2)= \frac{\sqrt{2}}{3} f_{\pi} \int_0^1 \mathrm{d} x \frac{\varphi_{\pi}(x,xQ)}{Q^2 x} 
\label{TFF}
\end{equation}
respectively. The pion Distribution Amplitude (DA), $\varphi_{\pi}$, appearing in Eq. \ref{TFF} is given by
\begin{align}
	\label{pionDA}
	f_{\pi} \varphi_{\pi}(x,\mu)=  2 \sqrt{\frac{N_c}{\pi}} \int \mathrm{d} b J_{0}(\mu b) b \{A((x(1-x) M_{\pi}^2) \nonumber \\+ B m_f M_{\pi}\} \frac{\Psi^{\pi} (x,\zeta)}{x(1-x)} \;.
\end{align}
A comparison of our spin-improved holographic DA with other DAs, including the platykurtic DA of \cite{Stefanis:2014yha} can be found in \cite{Ahmady:2016ufq}. Here, we focus on the following predictions: with $[A=0(1),B=1]$, we predict $\sqrt{\langle r_{\pi}^2 \rangle}=0.683(0.673)$ fm compared to $\sqrt{\langle r_{\pi}^2 \rangle}=0.544$ fm with $[A=1,B=0]$. The spin-improved predictions are thus in much better agreement with the experimental value: $\sqrt{\langle r_{\pi}^2 \rangle}_{\mathrm{exp}}=0.672 \pm 0.008$ fm \cite{Agashe:2014kda}. For the decay constant, we predict $f_{\pi}=135(138)$ MeV with $[A=0(1),B=1]$ compared to $f_{\pi}=161$ MeV with $[A=1,B=0]$. Comparing with the measured value, $f_{\pi}^{\mathrm{exp}}=130.4 \pm 0.04 \pm 0.02$ MeV \cite{Agashe:2014kda}, we again conclude that the spin-improved predictions are much favoured. Our predictions for the form factors are shown in Figures \ref{fig:EMFF} and \ref{fig:TFF}. As can be seen, the improvement in describing the data with the spin-improved wavefunctions are striking for both form factors at low $0 < Q^2 < 5~\mathrm{GeV}^2$, i.e. in the non-perturbative regime where the holographic light-front wavefunction is expected to be valid. Note that our pion DA, Eq. \ref{pionDA}, only evolves in the non-perturbative region, $0 \le \mu \le 1$ GeV, and thus lacks the known Efremov-Radyushkin-Brodsky-Lepage (ERBL) \cite{Lepage:1979zb,Efremov:1978rn} perturbative evolution for $\mu > 1$ GeV. 

A follow-up and extension of our work using (smaller) effective quark masses with a different ansatz for the pion spin structure can be found in \cite{Chang:2016ouf}.

 \section{Conclusions}
We have reported that it is possible to achieve a remarkable improvement in the simultaneous description of pion observables using a spin-improved light-front holographic wavefunction for the pion together with a universal AdS/QCD scale and without the need to invoke any important contribution of higher Fock sectors in the pion state.  Our predictions depend on the choice for the light quark masses and here we have used constituent masses to generate all predictions. 

\section{Acknowledgements}
This research is supported by NSERC Discovery Grants: SAPIN-2017-00031 (R.S) and SAPIN-2017-00033 (M.A). R.S thanks the organizers of QCD17 for a successful conference. 

\bibliographystyle{elsarticle-num}%
\bibliography{sandapen}%

\end{document}